\documentclass[pra,nofootinbib,twocolumn, superscriptaddress]{revtex4-2}
\usepackage{amsmath,amssymb,bm,graphicx}

\newcommand{\R}{\mathop{\mathbb{R}}\nolimits}

\newtheorem{Thm}{Theorem}
\newtheorem{Prop}{Proposition}

\newtheorem{Lem}{Lemma}

\begin{document}

\title{Relaxed Bell inequality as a trade-off relation between measurement dependence and hiddenness}

\author{Gen Kimura}
\email{gen@shibaura-it.ac.jp}
\affiliation{College of Systems Engineering and Science, Shibaura Institute of Technology, Saitama 330-8570, Japan}

\author{Yugo Susuki}
\affiliation{College of Systems Engineering and Science, Shibaura Institute of Technology, Saitama 330-8570, Japan}

\author{Kei Morisue}
\affiliation{Previous affiliation: Department of Physics, Waseda University, Tokyo 169-8555, Japan}

\begin{abstract}
Quantum correlations that violate the Bell inequality cannot be explained by any local hidden variable theory that is measurement independent. 
However, this violation merely signifies the incompatibility of the underlying assumptions of reality, locality, and measurement independence, without providing a quantitative measure of the extent to which each assumption is violated.
In contrast, Hall (2010, 2011) introduced measure of each assumption: indeterminism, signaling, and measurement dependence, and generalized the Bell-CHSH inequality that provides quantitative trade-off relationship between these assumptions.
In this paper, we consider the introduction of hidden variables to be an essential assumption of Bell's theorem and introduce a quantification of hidden variables, which we term ``hiddenness." We derive a new trade-off relation between hiddenness and measurement dependence that applies to any local hidden variable theory. 

\end{abstract}

\maketitle

\section{Introduction}

In 1964, Bell made a surprising discovery that quantum correlations cannot be explained by local realism \cite{ref:Bell,ref:Bell2}. 
He introduced an inequality satisfied by all local hidden variable models and showed that there are quantum correlations that violate the inequality when appropriately chosen set of measurements are made. 
The inequality is now known as the Bell inequality, and the sequence of results is referred to as Bell's theorem \cite{sep-bell-theorem}.
Importantly, the quantities appearing in the inequality are made up of measurable values (expected values or probabilities).  
As a result, these inequalities can be directly verified through experiments. 
Since the discovery of Bell's theorem, many experiments have confirmed (along with the verification of various loopholes) that certain quantum correlations violate the inequality as predicted by quantum theory \cite{PhysRevLett.28.938,PhysRevLett.49.1804,PhysRevLett.81.3563,Hensen2016,PhysRevLett.115.250402,PhysRevLett.115.250401,Gisin_Nicolas2014-07-30}. 
Consequently, it is widely believed that quantum theory is correct, and more importantly, that there indeed exists  phenomena that cannot be explained by local realism.
Subsequently, these quantum correlations have been found to be useful resources for various quantum information processings, such as quantum computer, quantum teleportation, and quantum cryptography \cite{ref:QI,ref:QI2}. 
Notably, the experimental fact of the violation of the Bell inequality is recognized as the key feature that makes some information processings to be completely secure, such as the key distribution \cite{PhysRevLett.67.661,PhysRevLett.97.120405,PhysRevLett.95.010503} and the random number generation \cite{Pironio2010,Colbeck2011}.  

In Bell's argument, however, the violation of Bell inequality tells us just that the underlying assumptions are logically incompatible. 
In particular, nothing can be said about how much each assumption must be violated to explain the experimental facts. 
In addition, clarifying the assumptions behind Bell's theorem is not so obvious as some of them are often implicit and can also vary between studies.
In Bell's original paper \cite{ref:Bell}, he explicitly assumed (i) determinism and (ii) locality in hidden variable models.
There is also another essential assumption that is often ``overlooked", namely (iii) that observers are free\footnote{Strictly speaking, we assume here that the measurement choice can be described by a probability theory. In contrast, it is also discussed how one can discuss a ``free will" that happens non-causally \cite{FreeWill} regardless of whether it can be described by a probability theory. } to choose measurements\cite{ref:MD1,Brans1988,Hall2010,PhysRevLett.106.100406}.  
Hence, Hall proposed quantitative measures for these three assumptions, referring to them as indeterminism ($I$), no-signaling ($S$), and measurement dependence ($M$), respectively.
Moreover, he successfully generalized the Bell-CHSH inequality \cite{ref:CHSH} that provides a trade-off relation between the empirical value $C$ (CHSH-value) and measures $I$, $S$, and $M$ in the simplest Bell scenario (with binary measurement settings and binary outcomes) \cite{PhysRevA.82.062117,PhysRevA.84.022102}.
To the best of the authors' knowledge, this is the first attempt to shed new light on Bell's theorem in this way.

Subsequently, several works appeared to follow this line of research. 
Another relaxed Bell inequality was derived for the models where only one party can violate measurement independence \cite{ref:Banik} and was further generalized in \cite{ref:Friedman} by introducing the measures for measurement dependence for each party. 
Recently, Ghadimi considered a model that relaxed the signaling property (usually referred to as parameter dependence \cite{ref:Jarret,ref:St} or active non-locality \cite{ref:Nelson}) and derived a relaxed Bell inequality \cite{ref:Ghadimi}. 

The significance of these studies extends to applications in informatics, where the underlying variables can be interpreted as potential resources for eavesdroppers.
Koh {\it et al.} showed a trade-off relation between CHSH value, the guessing probability, and the free will parameter (related with the measurement dependence introduced above) with the application of secure randomness expansion \cite{PhysRevLett.109.160404}. 
Another interesting relation was observed in \cite{Puetz2014} using the upper and lower bounds of the probability of measurement-contexts conditioned on the hidden variable for a measurement dependent local models. 

However, the very assumption of introducing hidden variables has never been subject to quantification thus far.
We claim that this assumption should also be quantitatively measured as one of the fundamental underlying assumptions in Bell's theorem. 
In this paper, we introduce such a measure by $H = \#(\Lambda)- 1$ where $\#(\Lambda)$ is the size (i.e., the cardinality) of the set of hidden variables $\Lambda$ and refer to $H$ as the ``hiddenness". 
If the variable is in the hands of the eavesdropper, the measure can be interpreted as the amount of information he/she can store.
We derive a new trade-off relation (Theorem \ref{thm:main}) between $H$ and measurement dependence $M$, which is satisfied for all (measurement dependent) local models \cite{Puetz2014}. 
In the case of $M=0$ (measurement-independent), the relation recovers the Bell-CHSH inequality. 
On the other hand, with the observed data that violates the Bell-CHSH inequality, $M$ and $H$ have a trade-off relation: To decrease $M$, $H$ must increase, and vice versa. 
Interestingly, the trade-off relationship saturates when $H \ge 3$ and coincides with Hall's inequality \cite{PhysRevA.84.022102} in the case of local models.
Therefore, this result is also a generalisation of Hall's inequality for local models.
We also show that our inequalities describe the necessary and sufficient condition of a local model, in the sense that not only all local models satisfy them, but also there is a local model with which $M$ and $H$ satisfy the inequalities.

To demonstrate these results, we technically derive the optimal CHSH value $C_{\rm opt}$ (see \eqref{eq:optCHSH}) that can be attained by some local model, and establish the lower and upper bounds of $C_{\rm opt}$ with given measurement dependence $M$ and hiddenness $H$ (Propositions \ref{Thm:LB} and \ref{Thm:LUB}). 
While the upper bound gives the new trade-off relation mentioned above, the lower bound also gives a quantitative estimate of the fact that the violation of Bell-CHSH inequality can still be explained by a local hidden variable model if we relax the measurement independence assumption. 

This paper is organized as follows: 
In Sec.~\ref{sec:HVM}, we provide a brief introduction of hidden variable theory and the measure for hiddenness.  
In Sec.~\ref{sec:RBI}, we show a relaxed Bell inequality as a trade-off relation with the measurement dependence and hiddenness. 
In Sec.~\ref{sec:Tightness}, we construct tight models that attain the equality. 
Finally, in Sec.~\ref{sec:CD}, we present our conclusion and discussion. 

\section{Measure for Hiddenness}\label{sec:HVM} 

Let us consider a bipartite physical system A and B (for Alice and Bob) where the measurements are supposed to be performed in spacelike separated regions. The experimentally accessible probability is the set of joint probabilities $p(a,b|x,y)$, where $x$ and $y$ denote the measurements performed by Alice and Bob, and $a$ and $b$ denote their respective outcomes. 
In the hidden variable theory, we introduce a hidden variable $\lambda \in \Lambda$ so that the empirical joint probabilities are obtained by averaging over the hidden variable:
\begin{equation}\label{eq:pdis}
	p(a,b|x,y) = \sum_{\lambda \in \Lambda} p(\lambda|x,y) p(a,b|x,y,\lambda), 
\end{equation}
where $p(\lambda|x,y)$ is the probability of $\lambda$ given the values of $(x,y)$, and $p(a,b|x,y,\lambda)$ represents the joint probability of the outcome $(a,b)$ given the values of $(x,y)$ and $\lambda$. 
Notice that one should replace the summation to the integral in the general situation. However, a non-trivial structure arises when we consider a model in which the size (i.e., cardinality) of the set $\Lambda$ of hidden variables is finite. Bearing this in mind, we use a summation symbol for the hidden variables, except in Appendix \ref{app:INF}, where we provide a proof for (continuously) infinite models.
In order to prove the Bell inequality, we need to assume both the locality condition and the measurement independence: The locality condition states that the outcomes $a$ and $b$ with fixed $\lambda$ are statistically independent, i.e., $p(a,b|x,y,\lambda) = p(a|x,\lambda) p(b|y,\lambda)$. 
(It's worth noting that the locality condition is equivalent to assuming both parameter independence and outcome independence \cite{ref:St,ref:Jarret,ref:Nelson}.) 
On the other hand, measurement independence asserts that the measurement context $(x,y)$ and $\lambda$ are independent: $p(\lambda|x,y) = p(\lambda)$. 
This condition can be interpreted as allowing us to choose the measurement contexts freely, even when the hidden variable is fixed \cite{ref:MD1,Brans1988,Hall2010,PhysRevLett.106.100406}. 
To summarize, in the context of local hidden variable theory with measurement independence, the following equation holds true:
\begin{equation}\label{eq:MDLHV}
	p(a,b|x,y) = \sum_{\lambda \in \Lambda} p(\lambda) p(a|x,\lambda) p(b|y,\lambda).  
\end{equation}
In this paper, we relax the measurement independence, and consider a (measurement dependent) local model \cite{Puetz2014}: 
\begin{equation}\label{eq:LM}
	p(a,b|x,y) = \sum_{\lambda \in \Lambda} p(\lambda|x,y) p(a|x,\lambda) p(b|y,\lambda). 
\end{equation}
Following the paper \cite{Hall2010}, we shall use the same measure for the measurement dependency:
\begin{equation}\label{eq:M}
	M := \sup_{x,y,x',y'} \sum_\lambda |p(\lambda|x,y) - p(\lambda|x',y')|.
\end{equation}
Note that $ 0 \le M \le 2$ and $M = 0$ if and only if the model is measurement independent. 
However, it is useful to express $M$ in terms of the total variation distance \cite{David_A_Levin2017-10-31} (sometimes known as the trace distance \cite{ref:QI} or the Kolmogorov distance \cite{ref:FG,Kimura2010}) as follows. 
The total variation distance between two probability measures $P, Q$ is defined by 
\begin{equation}\label{eq:DeltaE}
	\delta(P,Q) := \sup_{E} |P(E) - Q(E)|,
\end{equation}
where the supremum is taken over all the events $E\subset \Lambda$. 
In the discrete model $\Lambda =\{\lambda_1,\lambda_2,\cdots \}$, it is easy to see the relation $\delta(P,Q) = \frac{1}{2} 
\sum_\lambda |p_\lambda - q_\lambda|$ where $P,Q$ are given by the probability distributions $(p_\lambda)_{\lambda \in \Lambda}$ and $(q_\lambda)_{\lambda \in \Lambda}$, respectively. 
Therefore, we can express $M$ by  
\begin{equation}\label{eq:MTVD}
	M = 2 \sup_{x,y,x',y'}  \delta(P_{xy}, P_{x'y'}),
\end{equation}
where $P_{xy} = (p(\lambda|x,y))_{\lambda \in \Lambda}$ denotes the probability distribution for $\lambda$ with the measurement context $(x,y)$.  

As mentioned previously, the introduction of hidden variables is one of the essential assumptions underlying Bell's theorem. Therefore, we believe it is important to quantify this assumption in order to understand its meaning. 
In this paper, we introduce the following simple measure $H$, which we term {\it hiddenness}, defined by 
\begin{equation}\label{ref:Hid}
	H := \#(\Lambda) - 1 
\end{equation}
where $\#(\Lambda)$ is the cardinality of the set $\Lambda$. 
Obviously, $H \ge 0$ and only takes a discrete natural number (including infinity). 
One possible interpretation of this measure is how much we need to introduce a hidden variable to explain the empirical statistics. 
Another interpretation is the memory size available to the eavesdropper if the hidden variable is in their possession.

The minimum case, where $H=0$, corresponds to a trivial scenario where there is only one elementary event for the hidden variable. 
This essentially means there is no introduction of any hidden variable. 
The simplest but non-trivial case is $H=1$, where there are two hidden elementary events, $\Lambda = \{\lambda_1, \lambda_2\}$. 
This scenario is logically possible as one can imagine a world having a ``hidden coin" (e.g., $\lambda_1=$ ``tail" or $\lambda_2=$ ``head"). 
Similarly, for $H=2,3,\ldots$, we need $3,4,\ldots$ hidden elementary events for the hidden variable, respectively. 
The following section introduces a relaxed Bell inequality that establishes a trade-off relation between $M$ and $H$, which holds in all local models.


\section{Relaxed Bell inequality}\label{sec:RBI}

In this section, we examine the simplest Bell scenario, often referred to as the CHSH setting, which involves binary measurement settings $x,y= 0,1$ and binary measurement outcomes $a,b = \pm 1$.
As an experimentally accessible quantity, we consider the {\it CHSH value} defined by 
\begin{equation}\label{eq:CHSH}
	C := \langle 00 \rangle + \langle 01 \rangle + \langle 10 \rangle -\langle 11 \rangle, 
\end{equation}
where $\langle xy \rangle := \sum_{a,b= \pm 1} a b \ p(a,b| x,y) \ (x,y = 0,1)$ denotes the expectation value of the product of the measurement outcomes, for joint measurement setting $(x,y)$. 
It is well-known \cite{ref:CHSH} that for any measurement independent local hidden variable model \eqref{eq:MDLHV}, the CHSH value is always bounded from above by $2$, that is known as Bell-CHSH inequality:   
\begin{equation}\label{eq:BCHSH}
	C \le 2. 
\end{equation}
It should be emphasized that Bell-CHSH inequalities (the set of eight inequalities obtained by taking the absolute value and changing the position of the minus sign in \eqref{eq:CHSH}) provide not only a necessary condition but also a sufficient condition for the statistics to be explainable by measurement-independent local hidden variable models \cite{PhysRevLett.48.291}. This fact makes Bell-CHSH inequalities of special interest.

\begin{figure}[htb]
	\includegraphics[width=8cm]{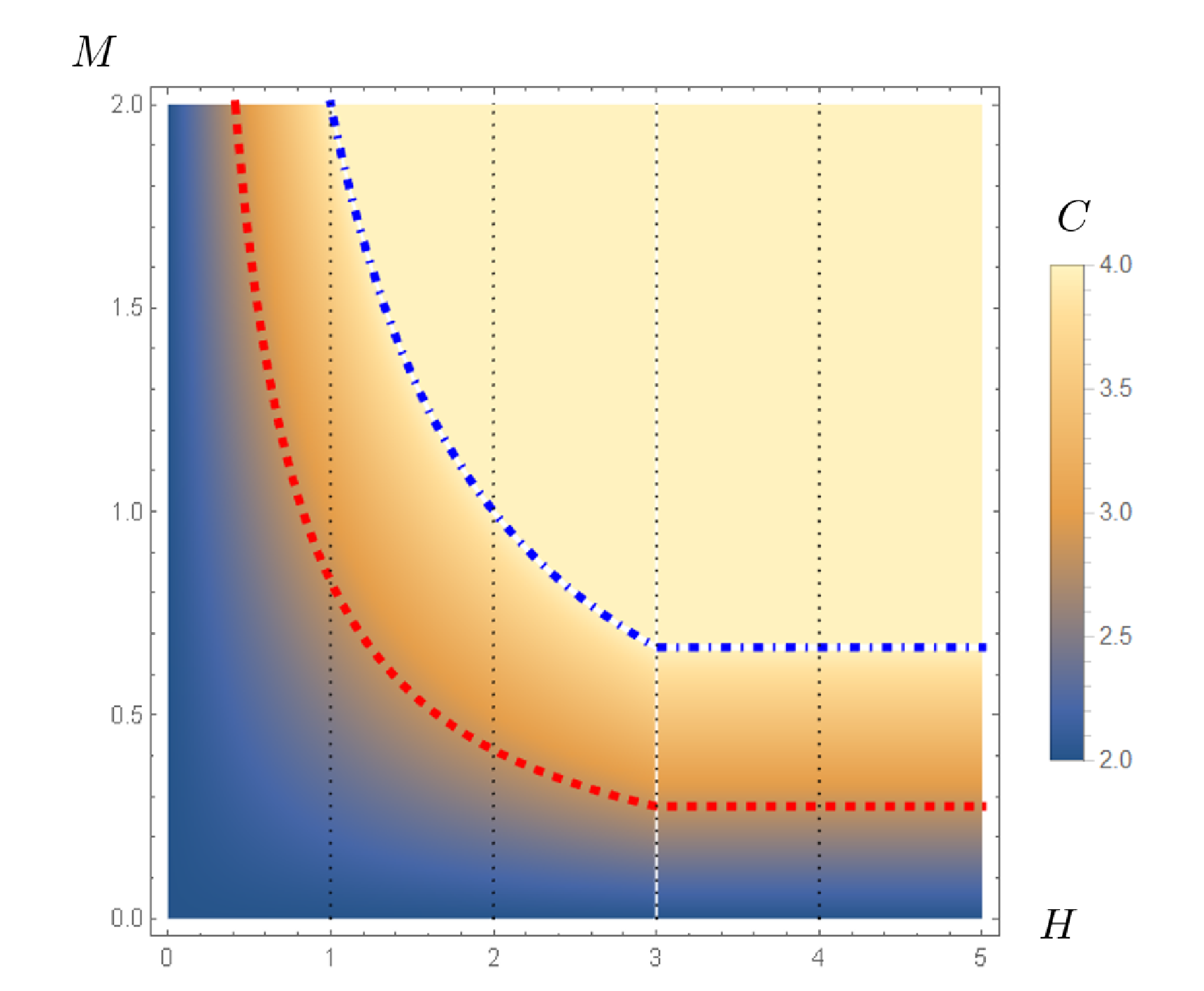}
	\caption{(color online) The upper bound of the CHSH Value $C$ in \eqref{eq:main} is plotted as a trade-off relation between $H$ and $M$. 
		The red dashed line corresponds to the maximal violation of the Bell-CHSH inequality in quantum systems: $C = 2 \sqrt{2}$. 
		The region above the blue dot-dashed line is a trivial violation: $C = 4$. 
	}\label{fig:TradeoffMH}
\end{figure}

Our main finding is the following: 
\begin{Thm}\label{thm:main}
For any local model, 
\begin{equation}\label{eq:main}
	C \le \min[H,3] M + 2,
\end{equation}
(as well as the trivial bound\footnote{Hence, the inequality can be written as $C \le \min \Bigl[\min[H,3] M + 2,4\Bigr]$. 
We adopt the form \eqref{eq:main} just to avoid this ugly expression. } $C \le 4$). 
The inequality is tight in the sense that there is a local model that can attains the equality. 
\end{Thm} 
Inequality \eqref{eq:main} is a generalization of Bell-CHSH inequality since it recovers the inequality by putting $M =0$ (measurement independence). 
In general, it provides a trade-off relation between CHSH value $C$, measurement dependence $M$ and hiddenness $H$. 
With a given violation of Bell-CHSH inequality ($C >2$), one can estimate the trade-off between $M$ and $H$: The less hiddenness $H$, the more measurement dependence $M$ is required, and vice versa (See Fig.~\ref{fig:TradeoffMH}). 
Suppose, for instance, that we observe the maximum violation of Bell-CHSH inequality in quantum theory, i.e., the Tsirelson bound $C = 2\sqrt{2} \simeq 2.8$. 
Then, for $H \ge 3$, one should give up the measurement independence at least $M = \frac{2}{3}(\sqrt{2}-1) \simeq 0.276$. 
For $H = 2$, $M$ should be greater than or equal to $\sqrt{2}-1 \simeq 0.414$ and, for $H = 1$, $M \ge 2(\sqrt{2}-1) \simeq 0.828$. 

An immediate corollary of Theorem \ref{thm:main} is that, for any local model,  
\begin{equation}\label{eq:Hge3}
		C \le 3 M + 2.
\end{equation}
This fact was previously observed by Hall \cite{PhysRevA.84.022102}, and therefore the relation \eqref{eq:main} generalizes his result for local models. 
Since this holds for any $H$, there exists an ultimate lower bound for $M$:
\begin{equation}\label{eq:SM}
	\frac{C-2}{3} \le M \ (\Leftrightarrow C \le 3M + 2).  
\end{equation}
Note that the case $H=0$, which corresponds to a local model essentially without the introduction of hidden variables, also gives the Bell-CHSH inequality. 

It would be interesting to consider the implications of our results for information security. 
For example, let us imagine that the hidden variable is in the possession of an eavesdropper, and that $H$ corresponds to the size of an exploitable information source. 
Our results suggest the following: In order to cheat legitimate users with an apparent violation of the Bell inequality based on a local model, the eavesdropper has to have a large control over the measurements of the legitimate users if the memory size $H$ is small. Conversely, if the eavesdropper has a large $H$, he/she does not need to worry much about controlling the measurements. Interestingly, however, there is a threshold ($H \geq 3$) beyond which the measurement dependence $M$ cannot be reduced any further.

In the following, we provide a proof of Theorem \ref{thm:main} in two steps: 
Firstly, we introduce ``the optimal CHSH value" $C_{\rm opt}$ that can be achieved by a local model. 
Secondly, we establish the tight upper bound (as well as the tight lower bound) for $C_{\rm opt}$, which yields \eqref{eq:main}. 
Although the proof is done with a discrete model, the result is still valid even for an uncountable model ($\#(\Lambda) = \infty$), which saturates to \eqref{eq:SM}. 
However, since the proof requires a slightly different approach for the uncountable case, it is presented in Appendix \ref{app:INF}.  

\bigskip 

\subsection{Optimal CHSH value for local model} 

Using local expectation values $A_x := \sum_{a} a p(a|x,\lambda) \ (x=0,1)$ and $B_y := \sum_{b} b p(b|y,\lambda) \ (y=0,1)$, the CHSH value \eqref{eq:CHSH} for any local model \eqref{eq:LM} is given by 
\begin{equation}\label{eq:CHSHalt}
    C = \sum_\lambda (z_{1} A_0 B_0 + z_{2} A_0 B_1 + z_{3} A_1 B_0 -z_{4} A_1B_1), 
\end{equation} 
where $z_i := p(\lambda|i)$ and $i=1,2,3,4$ corresponds to the measurement context $(x,y)= (0,0),(0,1),(1,0),(1,1)$ respectively. (In what follows, we will often use the same labeling $i$ for measurement contexts $(x,y)$ for convenience.) 
Notice here that all $\lambda$ dependencies in $A_x,B_y$ and $z_i$ are omitted.
We have a tight inequality:
\begin{equation}\label{eq:SSopt}
	C \le \sum_\lambda \Bigl(\sum_{i=1}^4 {z}_i - 2 \min_{i} {z}_i\Bigr)   
\end{equation}
from the following lemma:
\begin{Lem}\label{lem:f}
	For any positive tuple ${\bm z}:= (z_1, z_2,z_3,z_4) \in \R^{4}$, we have  
	\begin{eqnarray}
		\max_{
            \substack{
                A,A'\in [-1,1]\\
                B,B'\in [-1,1]
            }
            } 
            &&(
                 z_{1} A B + z_{2} A B' 
                 + z_{3} A' B -z_{4} A'B' 
		)\label{eq:maxeq}\\
		&&=g({\bm z}) := \sum_{i=1}^4 z_i - 2 \min_{i} z_i.
            \label{eq:gDef}
	\end{eqnarray}
\end{Lem}
(See Appendix \ref{sec:Proofs} for the proof.) 
Now we introduce ``the optimal CHSH value" for local models by:
    \begin{equation*}
        C_{\rm{opt}} :=\max_{p(a|x,\lambda), p(b|y,\lambda)}C. 
    \end{equation*}
By utilizing the normalization condition, i.e., $\sum_j \sum_\lambda p(\lambda|j) = \sum_j 1 = 4$ and \eqref{eq:SSopt}, we have an alternative form:
    \begin{equation}\label{eq:optCHSH}
    	C_{\rm{opt}} =\sum_{\lambda}g({\bm {z}})= 4 - 2 \sum_\lambda \min_{i} p(\lambda|i).  
    \end{equation}  
One observes the trivial inequality $C_{\rm opt} \le 4$ directly from this form.
It also can be shown that the equality is attained by an appropriate local probabilities $p(a|x,\lambda)$ and $p(b|y,\lambda)$ such that they attain the maximums in Lemma \ref{lem:f} for each $\lambda$.

In the subsequent subsections, we derive the lower and upper bounds of $C_{\rm opt}$ with given $H$ and $M$. 
The trivial case with $H = 0$ is described separately here. 
This case can be described by introducing a trivial set of hidden variable, i.e., a singleton set $\Lambda = \{\lambda_1\}$, so that 
\begin{eqnarray}
	p(a,b|x,y) = p(a|x,\lambda_1)p(b|y,\lambda_1),
\end{eqnarray}
and $p(\lambda_1 | x,y) = 1$ for all $x,y$. 
Since $M = 0$ is always satisfied in this case, the measurement dependence will not occur. 
We also have $ C_{\rm opt} := 4 - 2 \min_{i} p(\lambda_1|i) = 2$. 
Hence, only the case 
\begin{equation}\label{eq:trivialML}
	M = 0,\  C_{\rm opt} = 2 
\end{equation}
is possible for the trivial case $H = 0$. 

For non-trivial cases where $H= 1,2,3, \cdots$, measurement dependence $M$ can take any value in the range $[0,2]$. 
However, we will see below that a non-trivial lower and upper bounds of $C_{\rm opt}$ appear.

\subsection{Lower bound of $C_{\rm opt}$ for local model}

\begin{Prop}\label{Thm:LB}
	For any local model, 
	\begin{equation}\label{eq:LBofSopt}
		M + 2 \le C_{\rm opt}. 
	\end{equation}
	The inequality is tight. 
\end{Prop}
{\bf Proof.} We use the same notation $z_i := p(\lambda|i)$ and the labeling $i=1,2,3,4$ for $(x,y)= (0,0),(0,1),(1,0),(1,1)$ as in the previous subsection. 
Reminding the definition \eqref{eq:M}, let $i_1<i_2 \in \{1,2,3,4\}$ such that 
    \begin{equation*}
        M=\sum_\lambda |z_{i_1} - z_{i_2}|. 
    \end{equation*}
    Adding the normalization conditions $\sum_\lambda z_{i_3} = \sum_\lambda z_{i_4} = 1$ where $i_3 < i_4\in \{1,2,3,4\}\setminus \{i_1,i_2\}$, we observe  
    $M + 2 = \sum_\lambda (|z_{i_1} - z_{i_2}| + z_{i_3} + z_{i_4}) 
    \le \sum_\lambda \max [ (z_{i_1} - z_{i_2} + z_{i_3} + z_{i_4}),\  (-z_{i_1} + z_{i_2} + z_{i_3} + z_{i_4})]$.
    The last expression is bounded from above by $C_{\rm opt}$ since  $C_{\rm opt}=\sum_{\lambda}g({\bm z}) = \sum_{\lambda}\max[z_1+z_2+z_3-z_4,z_1+z_2-z_3+z_4,z_1-z_2+z_3+z_4,-z_1+z_2+z_3+z_4]$. 
 
Tightness of the inequality will be shown in Sec.~\ref{sec:Tightness}. \hfill $\blacksquare$ 

\bigskip 

    The proposition shows that for any given measurement dependence $M \in [0,2]$, there exists a local hidden variable model in which the CHSH value can reach $M+2$. 
    This reflects the often overlooked fact that a Bell inequality can be violated even by a local hidden variable model if measurement independence is relaxed \cite{ref:MD1,Brans1988,Hall2010,PhysRevLett.106.100406}. 

\subsection{Upper bound of $C_{\rm opt}$ for local model}

\begin{Prop}\label{Thm:LUB}
	For any local model,   
	\begin{equation}\label{eq:LbddM}
		C_{\rm opt}  \le \left\{
		\begin{array}{cc}
			2 & (H=0) \\
			M + 2 & (H=1) \\ 
			2M + 2 & (H=2) \\
			3 M + 2 & (H \ge 3)
		\end{array}
		\right.
	\end{equation}
	The inequalities are tight. 
\end{Prop}
Since there always exists a local model in which the CHSH value reaches $C_{\rm opt}$, this result proves Theorem \ref{thm:main}.

In particular, for $H=1$ ($\#(\Lambda)=2$) together with Proposition \ref{Thm:LB}, we have 
\begin{equation}\label{eq:LMc2}
	C_{\rm opt} = M + 2. 
\end{equation}
Namely, in this case, the optimal CHSH value and $M$ has an one-to-one relation (See Fig.~\ref{fig:SM} (a) ).

\bigskip 

{\bf [Proof for $H=0$]} We have already shown \eqref{eq:trivialML}, so the relation trivially holds. 

\bigskip 

For $H=1$, we provide a direct proof of \eqref{eq:LMc2} (i.e., \eqref{eq:LBofSopt} and \eqref{eq:LbddM} simultaneously):   

\bigskip 

{\bf [Proof for $H=1$]}
Let $\Lambda = \{\lambda_1,\lambda_2\}$. 
We denote for each $i=1,2,3,4$, ${z}_i := p(\lambda_1|i)$, so by normalization condition, $1-{z}_i = p(\lambda_2|i)$. 
Without loss of generality, we can assume ${z}_1 \ge {z}_2 \ge {z}_3 \ge {z}_4 \ge 0$. 
Noting that $|{z}_1-{z}_4|+|(1-{z}_1)-(1-{z}_4)| = 2({z}_1-{z}_4)$, etc., it is easy to see that $M = 2({z}_1-{z}_4)$.  
    Letting $w_i=\sum_{j=1}^4 {z}_j - 2 {z}_i \ (i=1,2,3,4)$ and invoking \eqref{eq:optCHSH}, we have 
    $
    C_{\rm opt} = 4-2(\min_i{z_i}+\min_{i}(1-z_i))=2-2{z}_4+2{z}_1. 
    $ 
Therefore, we have shown the equality $C_{\rm opt} = 2 + M$. 
\hfill $\square$

\bigskip 

In what follows, let $n = H- 1 = \#(\Lambda)\ (n \ge 3)$ and use a labelling for a hidden variable as $\Lambda = \{1,2,\ldots,n\}$ instead of writing $\Lambda = \{\lambda_1,\lambda_2,\ldots,\lambda_n\}$.

\bigskip 

{\bf [Proof for $H = 2,3$]}  
To prove the cases $H=2,3 \ (n=3,4)$, we use the following lemma:  
\begin{Lem}\label{lem:4=3} 
	For $n=3,4$, there exist $i_*,j_*= 1,2,3,4$ and $\lambda_* = 1,2,3$ such that 
	\begin{equation}\label{eq:lem4=3}
		\sum_{\lambda=1}^n p(\lambda|i_{\lambda}) + (n-1) |p(\lambda_*|i_*)-p(\lambda_*|j_*)| \ge 1,  
	\end{equation}
	where $i_\lambda \in \{1,2,3,4\}$ denotes an index which attains the minimum of $p(\lambda|i)$: 
	\begin{equation}\label{eq:MinIndex}
		p(\lambda|i_\lambda) = \min_{i} p(\lambda|i). 
	\end{equation}
\end{Lem}
(The proof of this lemma is given in Appendix \ref{sec:Proofs}.) 

By using \eqref{eq:MTVD}, \eqref{eq:optCHSH} and the notation \eqref{eq:MinIndex}, inequality \eqref{eq:LbddM} for $H = 2,3$ ($n = 3,4$) can be rewritten as 
    \begin{equation}\label{eq:MThm1}
    	\sum_{\lambda=1}^n p(\lambda|i_\lambda)  +  (n-1) \max_{i,j}\delta(P_{i},P_{j}) \ge 1  
    \end{equation}
    where $P_{i} = (p(\lambda|i))_{\lambda \in \Lambda}$. 
Let $i_*,j_*,\lambda_*$ be a pair with which \eqref{eq:lem4=3} is satisfied in Lemma \ref{lem:4=3}. 
Then, we have 
\begin{eqnarray}
	&&\sum_{\lambda=1}^n p(\lambda|i_{\lambda})   +  (n-1) \max_{i,j}\delta(P_{i},P_{j}) \nonumber \\
	&\ge& \sum_{\lambda=1}^n p(\lambda|i_{\lambda})  +  (n-1) \delta(P_{i_*},P_{j_*}) \nonumber \\
	&\ge& \sum_{\lambda=1}^n p(\lambda|i_{\lambda})  +  (n-1) |p(\lambda_*|i_*) - p(\lambda_*|j_*)| \ge 1 \label{eq:dtrick}. 
\end{eqnarray}
In particular, the second inequality follows by invoking the definition of the total variation distance in the form \eqref{eq:DeltaE} and applying the singleton event $E = \{\lambda_*\}$. \hfill $\square$ 

\bigskip 

The cases $H \ge 4$ will be proved by reducing the problems to the case $H = 3$ in the following manner: 

\bigskip 

{\bf [Proof for $H \ge 4$]} In a similar manner, what we have to show is, for $n \ge 5$,   
\begin{equation}\label{eq:n=5}	
	\sum_{\lambda=1}^n p(\lambda|i_{\lambda}) + 3 \max_{i,j} \delta(P_{i},P_{j}) \ge 1. 
\end{equation}
Define the partition of $\Lambda = \{1,2,\ldots,n\}$ by 
$$
E_\gamma = \{\lambda \in \Lambda \ | \ i_\lambda = \gamma\} \ (\gamma=1,\ldots,4).  
$$
(Remind the notation \eqref{eq:MinIndex}.) 
Clearly, we have $\cup_\gamma E_\gamma = \Lambda, \ E_\gamma \cap E_{\gamma'} = \emptyset \ (\gamma \neq \gamma')$. 
Some of $E_\gamma$ could be the empty set. 
Now, we introduce a new set of ``hidden variables" $\Gamma = \{\gamma \}_{\gamma= 1,2,3,4}$ with coarse-grained probabilities $\tilde{P}_i = (\tilde{p}(\gamma|i))_\gamma$ for each measurement context $i=1,\ldots,4$ where $\tilde{p}(\gamma|i) := P_i(E_\gamma) = \sum_{\lambda \in E_\gamma} p_i(\lambda|i)$.  
Following the notation \eqref{eq:MinIndex}, we denote by $\tilde{i}_\gamma \in \{1,2,3,4\} \ (\gamma \in \Gamma)$ with which it holds that $\tilde{p}(\gamma| \tilde{i}_\gamma) = \min_{i=1,2,3,4} \tilde{p}(\gamma| i)$. 
However, by the definition of $E_\gamma$, one can assume $\tilde{i}_\gamma = \gamma$ for all $\gamma$. 
To see this, one needs to show $\tilde{p}(\gamma|\gamma) \le \tilde{p}(\gamma|i)$ for any $\gamma$ and $i$. 
But, $\tilde{p}(\gamma|\gamma) = P_\gamma(E_\gamma) = \sum_{\lambda \in E_\gamma}  p(\lambda|\gamma)$.  
Since $i_\lambda = \gamma$ for any $\lambda \in E_\gamma$, one has $\sum_{\lambda \in E_\gamma}  
p(\lambda|\gamma) = \sum_{\lambda \in E_\gamma}  p(\lambda|i_\lambda) \le \sum_{\lambda \in E_\gamma}  p(\lambda|i) = P_i (E_\gamma) = \tilde{p}(\gamma|i)$. Applying Lemma \ref{lem:4=3} with $n=4$ for $\Gamma$ (noting $\#(\Gamma) = 4$), $\tilde{p}$ and $\tilde{i}_\gamma$, there exists $i_*,j_*,\gamma_* =1,\ldots,4$ such that 
\begin{equation}\label{eq:temp41}
	\sum_{\gamma=1}^4 \tilde{p}(\gamma|\tilde{i}_{\gamma}) + 3 |\tilde{p}(\gamma_*|i_*)-\tilde{p}(\gamma_*|j_*)| \ge 1.  
\end{equation}
Since $\tilde{i}_{\gamma} = \gamma$ and $i_{\lambda} =\gamma$ for $\lambda \in E_{\gamma}$, one has 
\begin{equation}\label{eq:temp42}
	\sum_{\gamma=1}^4 \tilde{p}(\gamma|\tilde{i}_{\gamma})=  \sum_{\gamma=1}^4 \sum_{\lambda \in E_{\gamma}} p(\lambda|\gamma) = \sum_{\lambda=1}^n p(\lambda|i_{\lambda})
\end{equation}
By applying the events $E_\gamma$ in the original definition of the total variation distance \eqref{eq:DeltaE}, we have 
\begin{eqnarray}\label{eq:temp43}
\max_{i,j} \delta(P_{i},P_{j}) &\ge& |P_{i_*}(E_{\gamma_*}) - P_{j_*} (E_{\gamma_*})| \nonumber \\
	&=& |\tilde{p}(\gamma_*|i_*)-\tilde{p}(\gamma_*|j_*)|
\end{eqnarray}
Combination of \eqref{eq:temp41}, \eqref{eq:temp42} and \eqref{eq:temp43} implies \eqref{eq:n=5}. 
\hfill $\square$

The tightnesses for all above cases will be shown in the next section.
\hfill $\blacksquare$

\section{Tight models}\label{sec:Tightness}
\begin{figure*}[t]
	\includegraphics[width=17cm]{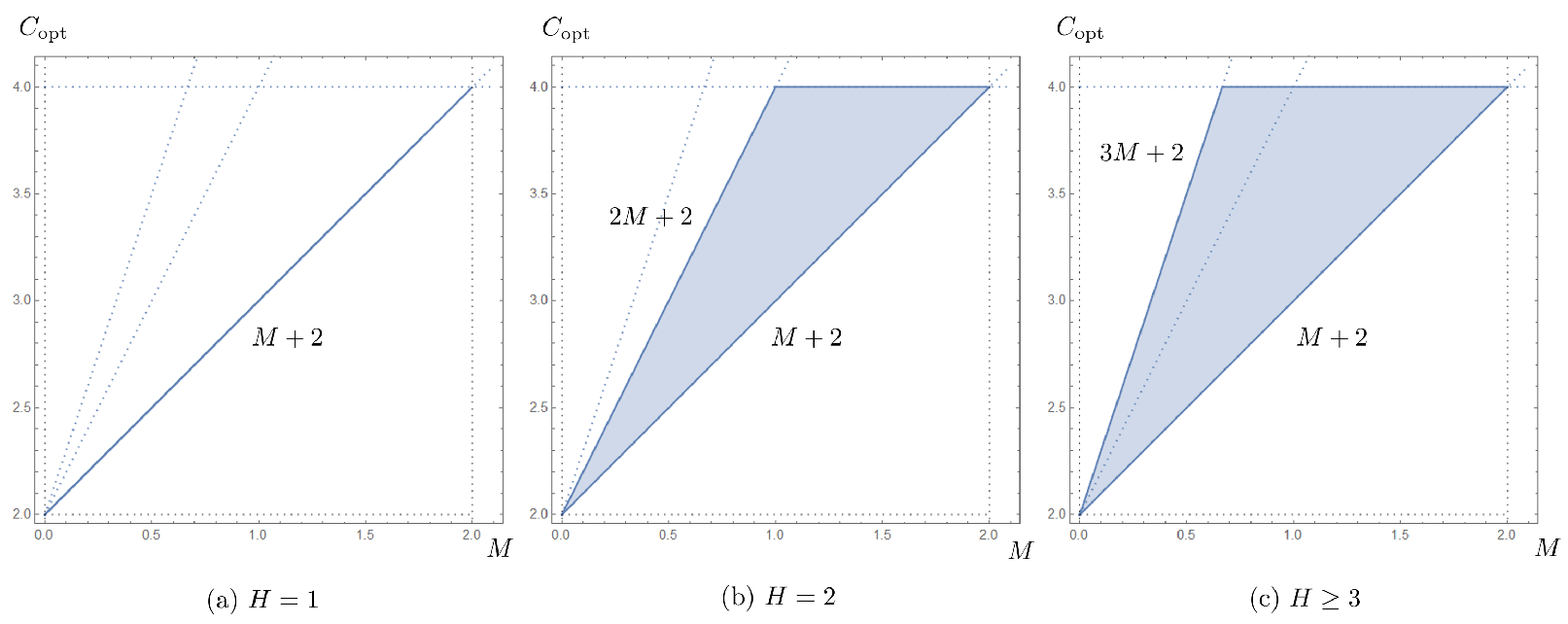}
	\caption{Relation between the measurement dependence $M$ and the optimal CHSH value $C_{\rm opt}$ for local model for $H=1,2$ and $H \ge 3$. The blue shaded regions are feasible regions by local models.}\label{fig:SM}
\end{figure*}

In this section, we demonstrate the tightnesses of inequalities in both Propositions \ref{Thm:LB} and \ref{Thm:LUB} by constructing explicit models of $p(\lambda|x,y)$ that achieve the equalities. 
Since there is a local model with some $p(a|x,\lambda)$ and $   p(b|y,\lambda)$ that achieves $C_{\rm opt}$, this demonstrates the tightness of Theorem \ref{thm:main}.

Note that the case $H=0$ trivially attains the bounds as shown in \eqref{eq:trivialML}. 
In the following, we show the tightnesses for the upper bounds in Proposition \ref{Thm:LUB} for the cases of $H=1,2$ and $H \ge 3$ in order. 

\bigskip 

[Case $H=1$] Let $\Lambda = \{\lambda_1,\lambda_2\}$. Let $P_{xy} = (p(\lambda|x,y))_{\lambda \in \Lambda}$ be given by Table \ref{tab:TMH1} with a one parameter $p \in [0,1]$. For this model, we have $C_{\rm opt} = 2p+2$ and $M = 2p$.  
Hence, $C_{\rm opt} = M + 2$ where $M$ runs over $[0,2]$ for $p \in [0,1]$. 

\bigskip 

[Case $H=2$] Let $\Lambda = \{\lambda_1,\lambda_2,\lambda_3\}$; Let $P_{xy} = (p(\lambda|x,y))_{\lambda \in \Lambda}$ be given by Table \ref{tab:TMH2} with a one parameter $p \in [0,1]$. For this model, we have $C_{\rm opt} = 4p + 2$ and $M = 2p$ for $p \in [0,1/2]$, hence $C_{\rm opt} = 2M + 2$ where $M$ runs over $[0,1]$ and $C_{\rm opt} = 4$ and $M = 2p$ for $p \in [1/2,1]$, hence $C_{\rm opt} = 4$ where $M$ runs over $[1,2]$.  

\bigskip 

[Case $H \ge 3$] Let $\Lambda = \{\lambda_1,\lambda_2,\lambda_3,\lambda_4,\ldots\}$; Let $P_{xy} = (p(\lambda|x,y))_{\lambda \in \Lambda}$ be given by Table \ref{tab:TMH3} with a one parameter $p \in [0,1]$. 
For this model, we have $C_{\rm opt} = 6p + 2$ and $M = 2p$ for $p \in [0,1/3]$, hence $C_{\rm opt} = 3M + 2$ where $M$ runs over $[0,2/3]$ and $C_{\rm opt} = 4$ and $M = 2p$ for $p \in [1/3,1]$, hence $C_{\rm opt} = 4$ where $M$ runs over $[1,2]$.

\begin{table}[h]
	\caption {Tight model for $H=1$ ($p \in [0,1]$)} \label{tab:TMH1}
	\begin{center}
		\begin{tabular}{c|cccc}
			$\lambda $ &  $P_{00}$ & $P_{01}$  & $P_{10}$  & $P_{11}$ \\ \hline \hline
			$\lambda_1$  & $0$ & $p$ &  $p$ & $p$ \\
			$\lambda_2$  & $1$ & $1-p$ & $1-p$ &$1-p$  \\
		\end{tabular}
	\end{center}
\end{table}

\begin{table}[ht]
	\caption {Tight model for $H=2$. (Left table for $p \in [0,1/2]$ and right table for $p \in [1/2,1]$)} \label{tab:TMH2}
	\begin{center}
		\begin{tabular}{c|cccc}
			$\lambda$ &   $P_{00}$ & $P_{01}$  & $P_{10}$  & $P_{11}$ \\ \hline \hline
			$\lambda_1$  & $0$ & $p$ &  $p$ & $p$ \\
			$\lambda_2$  & $p$ & $0$ & $p$ &$p$  \\
			$\lambda_3$  & $1-p$& $1-p$ & $1-2p$ &$1-2p$  \\
		\end{tabular}
		\begin{tabular}{c|cccc}
			$\lambda$ &   $P_{00}$ & $P_{01}$  & $P_{10}$  & $P_{11}$ \\ \hline \hline
			$\lambda_1$  & $0$ & $1-p$ &  $1-p$ & $2p-1$ \\
			$\lambda_2$  & $p$ & $0$ & $p$ &$1-p$  \\
			$\lambda_3$  & $1-p$& $p$ & $0$ &$1-p$  \\
		\end{tabular}
	\end{center}
\end{table}

\begin{table}[ht]
	\caption {Tight model for $H\ge 3$. (Left table for $p \in [0,1/3]$ and right table for $p \in [1/3,1]$)} \label{tab:TMH3}
	\begin{center}
		\begin{tabular}{c|cccc}
			$\lambda$ &   $P_{00}$ & $P_{01}$  & $P_{10}$  & $P_{11}$ \\ \hline \hline
			$\lambda_1$  & $0$ & $p$ &  $p$ & $p$ \\
			$\lambda_2$  & $p$ & $0$ & $p$ &$p$  \\
			$\lambda_3$  & $p$& $p$ & $0$ &$p$  \\
			$\lambda_4$  & $1-2p$& $1-2p$ & $1-2p$ &$1-3p$  \\
			$\lambda_5$  & $0$& $0$ & $0$ &$0$  \\
			$\vdots$ & $\vdots$& $\vdots$ & $\vdots$ &$\vdots$
		\end{tabular}
		\begin{tabular}{c|cccc}
			$\lambda$ &   $P_{00}$ & $P_{01}$  & $P_{10}$  & $P_{11}$ \\ \hline \hline
			$\lambda_1$  & $0$ & $\frac{1-p}{2}$ & $\frac{1-p}{2}$ & $p$ \\
			$\lambda_2$  & $p$ & $0$ & $\frac{1-p}{2}$ & $\frac{1-p}{2}$ \\
			$\lambda_3$  & $\frac{1-p}{2}$ & $p$ & $0$ & $\frac{1-p}{2}$ \\
			$\lambda_4$  & $\frac{1-p}{2}$ & $\frac{1-p}{2}$ & $p$ & $0$ \\
			$\lambda_5$  & $0$& $0$ & $0$ &$0$  \\
			$\vdots$ & $\vdots$& $\vdots$ & $\vdots$ &$\vdots$
		\end{tabular}
	\end{center}
\end{table}

Next, the lower bound in Proposition \ref{Thm:LB} can be attained by the model given in Table \ref{tab:TMH1} with trivial addition of $p(\lambda|x,y) = 0$ for $\lambda_i \ (i \ge 3)$.  

Moreover, one can show that the regions between lower and upper bounds of $C_{\rm opt}$ are feasible for any $M$ and $H$. 
Technically, this is far from trivial since both $M$ and $C_{\rm opt}$ are not affine function of $p(\lambda|x,y)$ in general. 
However, for probabilities given in Table \ref{tab:TMH1} (with trivial extension for any $H \ge 0$) for the lower bound and the ones given in Tables \ref{tab:TMH2} and \ref{tab:TMH3} for the upper bound, one can easily show that both $M$ and $C_{\rm opt}$ are affine for their convex combination. Hence, with these special choices of lower and upper bounds, their convex combinations fill the sandwiched regions. 
In Fig.~\ref{fig:SM}, the feasible regions for $M$ and $C_{\rm opt}$ given $H$ are shown by blue shaded regions.

\section{Conclusion and Discussion}\label{sec:CD}

In this paper, we introduced the measure of hiddenness $H$ and investigated a trade-off relation between $H$ and the measurement dependence $M$ for any local hidden variable models. 
In the CHSH setting, we derived a relaxed Bell inequality \eqref{eq:main} that generalizes the Bell-CHSH inequality. 
Interestingly, the structure of the trade-off changes between $H \le 2$ and $H \ge 3$: 
While the trade-off reduces to Hall's inequality \eqref{eq:Hge3} for $H \ge 3$, there appears a non-trivial dependence for $H$ when $H \le 2$. 
Moreover, the trade-off relation completely characterizes the range of measurement dependent local models.

In the present paper, hiddenness $H$ was introduced simply by the cardinality of the set of hidden variable, making $H$ a discrete quantity. 
In addition, this measure does not reflect the statistics of the hidden variables.
In the upcoming paper \cite{2022arXiv220813634T}, we overcome this disadvantage by introducing another measure of hiddenness that uses the max entropy. 
This new measure will provide a better reflection of the hidden variable statistics.
Furthermore, it would be interesting to generalize the results obtained in this paper by relaxing the condition of locality. 
To do this, we need to introduce measures for both parameter and outcome dependence.


\bigskip 

\section*{ACKNOWLEDGMENTS}
We are grateful to Michael Hall for his valuable comments and discussion. 
We also thank Ryo Takakura and Yuichiro Kitajima for their fruitful comments. 
G. K. is supported in part by JSPS KAKENHI Grants No. 17K18107.

\appendix

\section{Proofs of Lemmas}\label{sec:Proofs}

[Proof for Lemma \ref{lem:f}] Since the objective function $f_{\bm z}(A,A',B,B'):= z_{1} A B + z_{2} A B' + z_{3} A' B -z_{4} A'B'$ in \eqref{eq:maxeq} is affine for all variable $A,A',B,B'$, the maximum is attained by the extreme points $A,A',B,B' = \pm 1$. 
There are four cases to consider: [Case I] $A = A', B = B'$, [Case II] $A = A', B = - B'$, [Case III] $A = -A', B = B'$, [Case IV] $A = -A', B = - B'$. 
Direct computations show that [Case I] $f = \pm(z_1+z_2+z_3-z_4) = \pm (z - 2 z_4)$, [Case II] $f = \pm(z - 2 z_2)$, [Case III] $f = \pm(z - 2 z_3)$, [Case IV] $f = \pm(z - 2 z_1)$ where $z:= \sum_{j=1}^4 z_j$. 
This shows that 
\begin{equation}\label{eq:maxmax}
	\max_{A,A',B,B' \in [-1,1]} f_{\bm z}(A,A',B,B') =  \max_{i} |z - 2 z_{i}|.
\end{equation}
However, as is easily shown, the maximum is always attained by a positive $z - 2 z_{i}$. \hfill $\blacksquare$ 

\bigskip 

[Proof of Lemma \ref{lem:4=3}] We provide the proof for $n=3$. The case $n=4$ can be shown in parallel. 
Assume contrary for all $i,j \in \{1,2,3,4\}$ and $\lambda \in \{1,2,3\}$
\begin{equation}\label{eq:cont4=3}
	\sum_{\lambda'=1}^3 p(\lambda'|i_{\lambda'})  + 2 |p(\lambda|i)-p(\lambda|j)| < 1.  
\end{equation} 
There always exists $s \in \{1,2,3,4\}$ which is different from all $i_1,i_2,i_3$. 
Moreover, for each $\lambda$, one can choose $j_\lambda,k_\lambda \in \{1,2,3,4\}$ such that $i_\lambda,j_\lambda,k_\lambda,s$ are all different from each other, i.e., for any $\lambda$, $\{i_\lambda,j_\lambda,k_\lambda,s\} = \{1,2,3,4\}$. 

Applying the case $\lambda = 1$, $i=i_1,j=j_1$ to \eqref{eq:cont4=3}, one has 
$\sum_{\lambda'=1}^3 p(\lambda'|i_{\lambda'})  + 2 |p(1|i_1)-p(1|j_1)| = 
\sum_{\lambda'=1}^3 p(\lambda'|i_{\lambda'}) - 2 (p(1|i_1)-p(1|j_1))= 
-p(1|i_1) + p(2|i_2) + p(3|i_3)   + 2 p(1|j_1)$. Thus, one has 
$$
-p(1|i_1) + p(2|i_2) + p(3|i_3)   + 2 p(1|j_1) < 1 
$$
Also for the case $\lambda = 1$, $i=i_1,j=k_1$, one has 
$$
-p(1|i_1) + p(2|i_2) + p(3|i_3)   + 2 p(1|k_1) < 1.
$$
Similarly, for $\lambda = 2,3$, 
\begin{eqnarray*}
	p(1|i_1) - p(2|i_2) + p(3|i_3)   + 2 p(2|j_2) < 1,  \\
	p(1|i_1) - p(2|i_2) + p(3|i_3)   + 2 p(2|k_2) < 1,  \\
	p(1|i_1) + p(2|i_2) - p(3|i_3)   + 2 p(3|j_3) < 1,  \\
	p(1|i_1) + p(2|i_2) - p(3|i_3)   + 2 p(3|k_3) < 1.   \\
\end{eqnarray*}

Summing all above $6$ inequalities and dividing by $2$, one gets 
\begin{equation*}
    \begin{array}{ccc}
    p(1|i_1) & + p(1|j_1) & + p(1|k_1) \\
    + p(2|i_2) & + p(2|j_2) & + p(2|k_2)\\
    +  p(3|i_3) & + p(3|j_3) & + p(3|k_3)  
    \end{array}
      < 3.
\end{equation*}
Since $i_\lambda \neq j_\lambda \neq k_\lambda \neq s$ for all $\lambda \in \{1,2,3\}$, the left hand side can be grouped as 
$$
\sum_{i\neq s} \sum_{\lambda=1}^3 p(\lambda|i) = 3,
$$
which is contradictory and leads to the inequality \eqref{eq:cont4=3}.
\hfill $\blacksquare$



\section{Proof for infinite models}\label{app:INF}

In this appendix, we provide a proof of Theorem \ref{thm:main} for the case of an uncountable local hidden variable model with $\#(\Lambda) = \infty$, where the inequality \eqref{eq:main} reduces to inequality \eqref{eq:Hge3}. 
We need to replace \eqref{eq:LM} and \eqref{eq:M} with  
\begin{eqnarray}
    &&p(a,b|x,y) = \int d\lambda p(\lambda|x,y) p(a|x,\lambda) p(b|y,\lambda) \label{eq:LMInf}\\
    &&M := \sup_{x,y,x',y'} \int d\lambda \left| p(\lambda|x,y) - p(\lambda|x',y')\right |,\label{eq:MInf}
\end{eqnarray}
where $p(\lambda|x,y)$ is the probability density of $\lambda$ conditioned on the measurement context $(x,y)$.
 
The idea of the proof is to transform the uncountable hidden variable model with $\#(\Lambda) = \infty$ into another hidden variable model with $\#(\tilde{\Lambda})=2^4$: 
\begin{equation}
    \tilde{\Lambda}= \{\tilde{\lambda}=(a_0,a_1,b_0,b_1) \ | \ a_0,a_1,b_0,b_1 = 0,1\}.
\end{equation}
We introduce the probability distributions on $\tilde{\Lambda}$ by 
\begin{widetext}
        \begin{equation}\label{eq:q}
            q(a_0,a_1,b_0,b_1| x,y) := \int d\lambda p(\lambda|x,y) p(a_0|x=0,\lambda) p(a_1|x=1,\lambda) p(b_0|y=0,\lambda)p(b_1|y=1,\lambda)	
        \end{equation}
        and a deterministic model given by:
        \begin{equation*}
            q(a,b|x,y) := \sum_{a_0,a_1,b_0,b_1=0,1} q(a_0,a_1,b_0,b_1| x,y) \delta_{a a_x}\delta_{b b_y}.
        \end{equation*}
        It can be easily shown that $p(a,b|x,y) = q(a,b |x,y)$, and hence the CHSH value does not change:  
        \begin{equation*}
            C=\tilde{C}.
        \end{equation*}
    Let $\tilde{M}$ be the measurement dependence for this new hidden variable model on $\tilde{\Lambda}$: 
    \begin{equation}\label{eq:NewM}
        \tilde{M} := \sup_{x,y,x',y'} \sum_{a_0,a_1,b_0,b_1=0,1}| q(a_0,a_1,b_0,b_1| x,y) - q(a_0,a_1,b_0,b_1| x',y')|
    \end{equation} 
    By substituting \eqref{eq:q} into \eqref{eq:NewM}, we have 
    \begin{eqnarray*}
        \tilde{M} &=& \sup_{x,y,x',y'} \sum_{a_0,a_1,b_0,b_1=0,1}\Bigl|  \int d\lambda \Bigl(p(\lambda|x,y)-p(\lambda|x',y')\Bigl) p(a_0|x=0,\lambda) p(a_1|x=1,\lambda) p(b_0|y=0,\lambda)p(b_1|y=1,\lambda)\Bigr| \\
        &\le & \sup_{x,y,x',y'} \sum_{a_0,a_1,b_0,b_1=0,1} \int d\lambda \Bigl|p(\lambda|x,y)-p(\lambda|x',y')\Bigl| p(a_0|x=0,\lambda) p(a_1|x=1,\lambda) p(b_0|y=0,\lambda)p(b_1|y=1,\lambda) \\
            &= & \sup_{X,Y,X',Y'} \int d\lambda \Bigl|p(\lambda|x,y)-p(\lambda|x',y')\Bigl| \\
         &=& M,
    \end{eqnarray*}
where we have used the triangle inequality for the integral and the normalization conditions for $p(a|x,\lambda)$ and $p(b|y,\lambda)$.
        
Since the model on $\tilde{\Lambda}$ is finite, we have already shown that 
        $$
        C \le 3 \tilde{M} + 2. 
        $$
However, as shown above, we have $\tilde{M} \le M$, this completes the proof of \eqref{eq:Hge3} in the uncountable model. 
\hfill $\blacksquare$
\end{widetext}


\bibliography{ref_RelBell}

\end{document}